\documentclass[12pt]{iopart}

\expandafter\let\csname equation*\endcsname\relax
\expandafter\let\csname endequation*\endcsname\relax
\usepackage{amsmath}
\usepackage{amssymb}
\usepackage{mathtools}
\usepackage{iopams}
\usepackage{xcolor}

\begin{document}

\title[Combined effect of incentives and coupling in multigames]{Combined effect of incentives and coupling in multigames in two-layer networks}

\author{Luo-Luo Jiang$^{1,*}$, Yi-Ming Li$^{1}$, Wen-Jing Li$^{2}$ and Attila Szolnoki$^{3,*}$}
\address{$1$ School of Information Technology and Artificial Intelligence, Zhejiang University of Finance and Economics, Hangzhou, 310018, China\\
$2$ College of Information Engineering, Zhejiang University of Water Resources and Electric Power, Hangzhou, Zhejiang 310018, China\\
$3$ Institute of Technical Physics and Materials Science, Centre for Energy Research, P.O. Box 49, H-1525 Budapest, Hungary\\
$*$ Corresponding author}
\ead{\\jiangluoluo@zufe.edu.cn (L.-L. J.)\\ szolnoki.attila@ek-cer.hu (A. S.)}

\vspace{10pt}
\begin{indented}
\item[]
\end{indented}

\begin{abstract}
The lack of cooperation can easily result in inequality among members of a society, which provides an increasing gap between individual incomes. To tackle this issue, we introduce an incentive mechanism based on individual strategies and incomes, wherein a portion of the income from defectors is allocated to reward low-income cooperators, aiming to enhance cooperation by improving the equitable distribution of wealth across the entire population. Moreover, previous research has typically employed network structures or game mechanisms characterized by homogeneity. In this study, we present a network framework that more accurately reflects real-world conditions, where agents are engaged in multiple games, including prisoner's dilemma games in the top-layer and public good games in the down-layer networks. Within this framework, we introduce the concept of ``external coupling'' which connects agents across different networks as acquaintances, thereby facilitating access to shared datasets. Our results indicate that the combined positive effects of external coupling and incentive mechanism lead to optimal cooperation rates and lower Gini coefficients, demonstrating a negative correlation between cooperation and inequality. From a micro-level perspective, this phenomenon primarily arises from the regular network, whereas suboptimal outcomes are observed within the scale-free network. These observations help to give a deeper insight into the interplay between cooperation and wealth disparity in evolutionary games in large populations.

\end{abstract}

\noindent{\it Keywords}: social dilemmas, cooperation, multigames, multi-layer networks

\section{Introduction}\label{intro}
In recent years, the global challenge of greenhouse gas emissions has drawn parallels to the tragedy of the commons, as first warned by Hardin in 1968~\cite{HardinS1968}. Despite the prevalence of cooperative behaviors across various domains, insufficient joint efforts still hold the potential for failure in averting certain disasters. Therefore, it is crucial to delve into the drivers of cooperation and strategies for enhancing cooperative processes~\cite{CelikerTCB2013,MitaniAB2000,SzaboPR2007,InmanNCC2009}. The emergence of cooperative behavior in human societies has been studied by evolutionary game theory~\cite{HamiltonJTB1964,NowakS2006}. Consequently, several game types have been suggested to mirror realistic conflicts, such as the prisoner's dilemma game (PDG)~\cite{rapoport}, public goods game (PGG)~\cite{perc13}, spatial snowdrift game~\cite{ZhangAMC2023}, boxed pigs game~\cite{WangPRE2021}, and multigames~\cite{WangPRE14,DengCSF2018,szolnoki14}. As a summation of early-stage research efforts, Nowak outlined five categories of cooperation-promoting mechanisms in response to the fundamental features of cooperative behavior among humans: Kin Selection, Direct Reciprocity, Indirect Reciprocity, Network Reciprocity, and Group Selection~\cite{NowakS2006}. Subsequent studies have identified additional potential cooperation-promoting mechanisms such as environmental feedback~\cite{MaAMC2023,KleshninaNC2023,szolnoki17}, time delay~\cite{PalCIJNC2020,YanNJP2021}, decreasing group size~\cite{JiangAMC2021}, long-term benefits~\cite{DankuSR2019}, aspiration~\cite{LiuPRE2016,PalABC2021}, etc. Together, these investigations illuminate the diverse array of factors influencing changes in the dynamics of collective cooperation.

The concept of network reciprocity, first observed by Nowak and May~\cite{NowakN1992}, has sparked extensive research into the evolution of cooperative behavior on different network structures that can faithfully model the structure of real populations~\cite{NowakS2006,SzaboPRE2009,WangJTB2014}. In addition to studying the evolution of cooperation on static networks, such as regular networks and highly heterogeneous scale-free networks~\cite{PoncelaPRE2011,XuPRE2017,ZhouCIJNS2021}, researchers have also focused on dynamic networks, including temporal networks~\cite{LiNC2020}, adaptive networks~\cite{ZschalerNJP2010,ChengNJP2011}, and networks generated by using the Watts-Strogatz algorithm~\cite{ChattopadhyayCIJNS2020}. Within the framework of network reciprocity, incentive mechanisms have been incorporated to support cooperators and to discourage defective or free-riding behaviors through the implementation of rewards and punishments, respectively~\cite{SzolnokiEPL2010,SWangAMC2022}. However, the application of these incentive mechanisms often implies a high cost, especially as reward is typically more expensive than punishment, potentially jeopardizing the sustainability of such mechanisms. Furthermore, Zhou {\it et al.} have emphasized the emergence of wealth disparities resulting from incentive mechanisms, indicating that these mechanisms only maximize total wealth when the enforcer's wealth closely aligns with that of others in the network~\cite{DuongPRSA2021,FloresAMC2024,ZhouPRSMPES2022}. Currently, it remains an open question to identify those potential factors that influence the effectiveness of incentives. Further research should aim to assess the impact of incentives on cooperation and wealth among individuals with diverse characteristics in a manner that better mirrors real-world conditions.

A notable trend in recent studies is the emphasis on multi-layer networks, just as solving the issue of greenhouse gas emissions requires a multinational effort at different levels~\cite{InmanNCC2009}. The coupling strength between multi-layer networks or the connection strength between interdependent networks has been proven a facilitating effect on cooperation~\cite{LiNJP2010,SuAMC2023,ChenAMC2021,JiangSR2015}. Moreover, the asymmetry of structures of multi-layer networks can lead to variations in cooperation because of the diverse social relations between individuals in many scenarios~\cite{WangPRE2021,W.-J. LiND2024}. It is essential to consider the variety of social dilemmas encountered by individuals across these distinct networks. This assumption offers a more accurate depiction of real-world interactions, as each layer signifies a specific relationship among participants, which may differ significantly from those represented in another layer.

It is also a crucial point that resources are limited in real-life scenarios. Furthermore, the unequal distribution of resources contributes to wealth disparities among individuals, which can have either positive or negative impacts on cooperation. Li {\it et al.} have pointed out that inequality between economic units under the climate commons dilemma serves to alleviate the social dilemma of climate warming~\cite{LiCC2021}. At the same time, in other papers, using complex networks, economic inequality has been found to promote the emergence of social cooperation, but there are normally restrictions, such as a substantial reduction in resources or only a moderate level of inequality~\cite{AhmedPRE2014,WangAMC2022,ChenAMC2023}. However, individuals tend to cooperate more with wealthier counterparts in the presence of indulged wealth disparities, and network rewriting exacerbates such inequality. Even severer inequality has additional adverse effects, including hinder economic freedom, diminishing individual happiness, and accelerating the decline of cooperation~\cite{MelamedSR2022,IslamJCE2018,SkewesCRESP2023}. The majority of previous studies have primarily explored the impact of wealth disparities in terms of the initial endowment of individual wealth, while there has been limited research on how to alleviate wealth disparities during the process of games.

Two crucial issues remain unexplored in the aforementioned studies. Firstly, although previous studies have focused on the two themes of two-layer networks and incentive mechanisms, just a few have investigated them together. Notably, this aspect is important for the role of inter-organizations in the extensive information dissemination presence. This paper addresses this gap by investigating the combined effects of inter-network coupling strength and incentive strength on cooperation. A counterintuitive argument made by Granovetter has been embraced by social sciences, i.e., individuals are more likely to prefer weak social ties with acquaintances for new information than strong ties with close neighbors~\cite{Granovetter1977}. The research groups of Park and Jahani have supported this argument, finding that long-ties relationships such as ``neighbors of neighbors'' affect tie strength and are robustly associated with economic prosperity~\cite{JahaniPNAS2023,ParkS2018}. These papers inspired the introduction of ``external coupling'' to two-layer networks under a multigames setting. External coupling ``invisibly'' connects two individuals on different networks into acquaintances, and individuals have access to the benefits datasets of the corresponding network acquaintances and ``acquaintances of acquaintances''. At the same time, incentives should prioritize the participation and self-governance of local communities, as external control measures may undermine the ability of communities to manage shared resources autonomously, leading to overexploitation or unsustainable use of resources~\cite{OstmannR1998}. A prosocial organization facilitates cooperation by charging very little in the way of benefits, but only if it is sufficiently forward-looking~\cite{Chiba-Okabe2024}. By controlling for less intense incentives, the incentive setup in this paper allows for the concept of incurred costs to be ignored, thus providing a fresh perspective on the positive steps that prosocial institutions might take. The incentive mechanism also considers individual payoff as an incentive perspective, ultimately aiming at mitigate the widening disparity in wealth through redistribution. Secondly, existing research lacks an explanation of how wealth disparities over the course of social evolution can be reduced. This paper complements the existing literature by providing insights into the changes in wealth disparities caused by mechanisms and reflected by the Gini coefficient. Furthermore, we offer a microscopic perspective on the changes engendered by the mechanism on individuals within separate layers of the network, thereby enhancing the persuasiveness of our conclusions. Our results demonstrate the effectiveness of this mechanism. The incentive mechanism enlarges the cooperation on the two-layer network, and the external coupling strength within a certain range also has a positive consequence. Moreover, wealth disparities are strongly correlated with human interactions, with higher rates of cooperation corresponding to lower Gini coefficients.

The structure of this paper is as follows. We first present the implementation of the two-layer network model and evolutionary dynamics in Section~2. The main simulation results are discussed in Section~3. Finally, our conclusions are given in Section~4.

\section{Model}

We propose a new framework of a two-layer network that combines two different topological network structures, where agents participate in multiple games and are constrained by an incentive mechanism aimed at preventing significant wealth disparities. This mechanism combines the strategies adopted by agents with the resulting returns, which is closer to the decision-making process of individuals in real society. There are both positive and negative incentives, inducing agents to receive rewards or penalties. This leads to a learning process, where agents are more likely to emulate the strategies of those who have achieved greater success through these positive or negative incentives. Therefore, our model consists of two main components: the network structure and the evolutionary dynamics, as introduced below. The latter can be broken down into three stages: the multigames stage, the wealth redistribution stage under the incentive mechanism, and the strategy updating process. Through this comprehensive approach, we aim to better understand the overall effects of two-layer network structures and incentive mechanisms on agent behavior and outcomes in a multigame environment.

\begin{figure}[h!]
	\centering
    \includegraphics[width=0.4\textwidth]{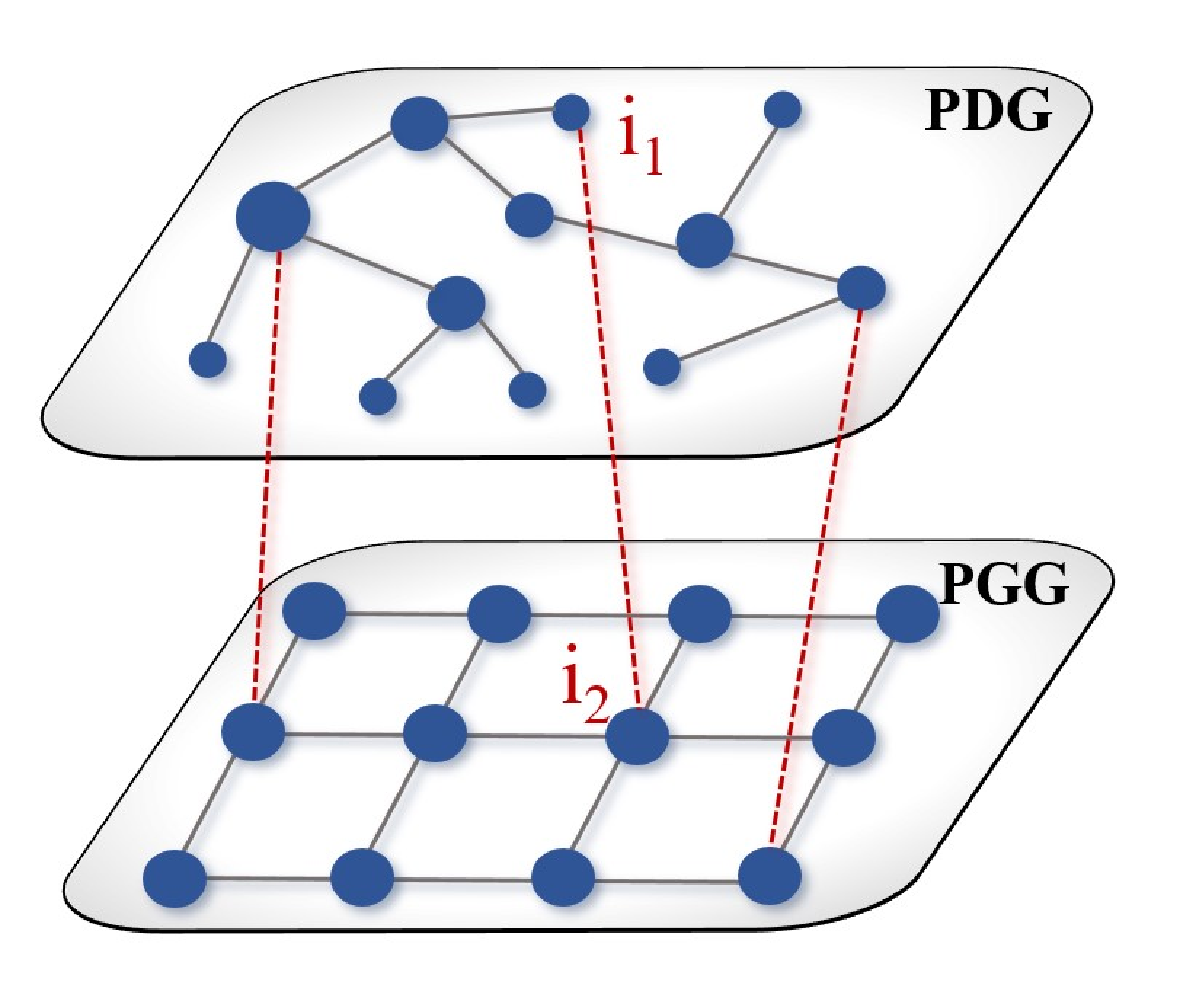}
	\caption{Topological structure of our model. It involves a two-layer network with external coupling (red dashed lines), where the upper layer is a scale-free network and the lower layer is a regular network. Each node represents an agent where the size is proportional to its degree (the number of neighbors). A solid line between two nodes indicates the neighbor relationship. All lines are undirected.
		\label{Fig:fig1} }
\end{figure}

\subsection{Network structure}
A two-layer network is established, which consists of an irregular scale-free network and a regular network with an equal number of agents, as shown in Fig.~\ref{Fig:fig1}. The connection details between the two-layer network are as follows. An agent $i_1$ who is traversed on average in the scale-free network chooses with probability $p$ ($ 0\leq p\leq1 $) whether to establish a connection with another agent $i_2$ in the regular network, which is restricted to only one per agent. If $i_1$ does, it captures access to a benefits dataset containing benefits information from $i_2$ and $i_2$'s neighbors. These cross-network connections are termed ``external coupling'', representing ``invisible'' links between agents not physically connected otherwise. Here, $p$ is referred to as an ``external coupling strength''. Consequently, external coupling facilitates the exploration of relations between groups and the analysis of multilayered social structures~\cite{Granovetter1977}. If $ p=0 $, the two network layers are completely independent, and strategy updating occurs solely within each single-layer network. By contrast, if $p=1$, every agent possesses an external coupling, where their strategies are fully influenced by those of agents in the other network. This framework enables us to explore the impact of agents' external couplings on cooperation and disparities of payoffs among distinct socioeconomic groups.

\subsection{Multigames model}
The applied multigames consist of the Prisoner's Dilemma Game and the Public Goods Game. In the initial phase of the game, agents participating in these games are defined as either cooperators (C) or defectors (D) with a uniform probability, based on their respective layers within the two-layer network. Agents play the PDG in the scale-free network. When both two agents choose to cooperate (defect), they will receive 1 (0). When a cooperator interacts with a defector, the former receives 0 while the latter receives 1.2~\cite{NowakN1992}. On the other hand, agents play the PGG in the regular network, which presents a complex interplay between individual decisions and network dynamics. Agents participate in $ G=k_{i_2}+1 $ groups gaming centered on themselves and their neighbors, where the parameter $k_{i_2}$ represents the number of neighbors of $i_2$ ($k_{i_2}$ is constantly 4 in the applied regular network). In a PGG, the aggregate contribution to the public pool is the sum of the costs ($c$) invested by each agent, with $ c=1 $ if the agent is a cooperator or $ c=0 $ if the agent is a defector. When $i_2$ participates in a PGG with its neighbor $j_2$ as the focal participant, the public pool is multiplied by the synergy factor ($r$) and is divided equally among all $G=5$ members. Furthermore, player $i_2$'s gain is then reduced by its personal cost. Therefore, the synergy factor and neighborhood structure play crucial roles in shaping the evolution of cooperation. The total payoff for $i_2$ is accumulated gains originating from all involved groups, which is expressed as:
\begin{equation}
	\pi_{i_2} = \sum_{j_2=0}^{\Omega_{j_2}}\pi_{i_2j_2}\,.
\end{equation}

\subsection{Incentive mechanism}
Similar to the real world, agents employing different strategies experienced alterations in their actual wealth level. The implementation of the incentive mechanism is conducted separately in each network with progressive steps for redistributing wealth, as the games played in the two networks are distinct, which leads to overall wealth varying from one network to another. Primarily, we classify individuals in a network as high-income or low-income by comparing their payoffs with the network's average. Those with payoffs above the average are considered high-income, while those below are classified as low-income. In this way, we can analyze wealth distribution in the network by enhancing payoff differences, leading to insights into income dynamics and disparities for potential intervention. According to our incentive mechanism, there will be four attribute groups of the population including high-income cooperators (HC), high-income defectors (HD), low-income cooperators (LC), and low-income defectors (LD). To quantitatively depict the severity of the penalization imposed on a defector, or to determine the amount of funds available for redistribution, we introduce the parameter $\alpha$ ($ 0<\alpha<1 $).

The redistribution incentive pool of reward premiums is:
\begin{equation}
	I = \sum_{i_1=0}^{\Omega_{HD}}\alpha\pi_{i_1}+\sum_{i_1=0}^{\Omega_{LD}}\alpha\pi_{i_1}
\end{equation}
The ultimate reward premiums are evenly distributed among each LC player. Upon conclusion of the incentive phase, the final benefit for each agent is as follows:
\begin{equation}
	\Pi_{i_1}=\begin{cases}(1-\alpha)\pi_{i_1},\quad &\forall i_1\in(\Omega_{HD}\cup\Omega_{LD})\\I/N_{LC}+\pi_{i_1},\quad &\forall i_1\in\Omega_{LC}\\\pi_{i_1},\quad &\forall i_1\in\Omega_{HC}\,.\end{cases}
\end{equation}
Here $\Omega_{HC}$, $\Omega_{HD}$, $\Omega_{LC}$ and $\Omega_{LD}$ represent the sets of HC, HD, LC, and LD within only one network, respectively. $N_{LC}$ represents the total number of LC within the same network.

The penalty structure is proportional, meaning that wealthier defectors incur more significant penalties. If $ \alpha=0 $, there is no additional incentive for low-income cooperators. If $ \alpha=1 $, all the payoffs received by the defectors are fully forfeited to the redistribution incentive pool. Due to the regular-free environment of the incentive setting, this paper should control $ \alpha $ in a smaller interval to more closely represent the real-world implications. Despite the societal inevitabilities of inequality, our objective is to mitigate the disparities between low-income cooperators and defectors, especially among high-income defectors, in such an environment that ignores the concept of cost. Our intention is for this mechanism to not only address unfairness towards disadvantaged groups and improve general well-being but also ensure the long-term sustainability of cooperative behavior in society. Therefore, we focused our research on conditions of small values of $\alpha$ ($\alpha < 0.3$), which means the cost of punishment is neglected.

\subsection{Strategy updating}
During the strategy learning phase, when the incentive is already applied, each of the agents within the two-layer network has an opportunity to make a strategy change. There are two different ways to update strategy depending on whether an agent has external coupling or not. In the first case, when $i_1$ has an external coupling, marked by a red dashed line in Fig.~\ref{Fig:fig1}, the strategy choice will be influenced only by the down-layer network. The specific method is that $i_1$ has full knowledge about $i_2$ and $i_2$'s neighbors, and will adopt the strategy of the most successful agent from the mentioned set. In an alternative way, when $i_1$ does not have external coupling, it will choose a random neighbor $j_1$ within the upper layer network, and compare the pros and cons of the benefit of $j_1$ to choose whether or not to learn its strategy. Importantly, it is not a deterministic decision, and its uncertainty is implemented by using the so-called Fermi's function to formulate the probability of $i_1$ learning from $j_1$:
\begin{equation}
	q=\frac{1}{1+\exp[(\Pi_{i_1}-\Pi_{j_1})/K]}\,
\end{equation}
Where $K>0$ reflects the level of noise in the learning process. In the $K \to 0$ limit agents are fully rational with no external interference, and while in the $ K\rightarrow\infty $ limit learning is completely randomized. In agreement with several previous works, we choose the fixed $K=0.1$ level in this paper. The strategy updating process of $i_2$ is the same and synchronized with $i_1$, i.e., $i_2$ also adopts the strategy from $i_1$ and $i_1$' neighbors or uses Fermi's function to update.

We introduce the cooperation rate $ f_C=N_C/(N_C+N_D) $, a metric employed to observe the progression of cooperation, which gauges the extent of cooperative behavior within networks. Here, $N_C$ and $N_D$  denote the total number of cooperators and defectors within the two-layer network. This study conducts a series of simulation experiments using the devised model. As a start, each independent experiment is based on a diverse two-layer network with $ N=4096 $ ($L=64$) nodes per layer for the evolutionary game. In the next step, the simulation experiments for each parameter undergo 10,000 iterations, and following a sufficient duration, the averages of the last 1,000 iterations are calculated to ensure the stability of the evolutionary process. To reach the expected accuracy, we averaged the results over 50 independent experiments.

\section{Simulation results}\label{sec:results}
\begin{figure*}
	\centering
    \includegraphics[width=.8\textwidth]{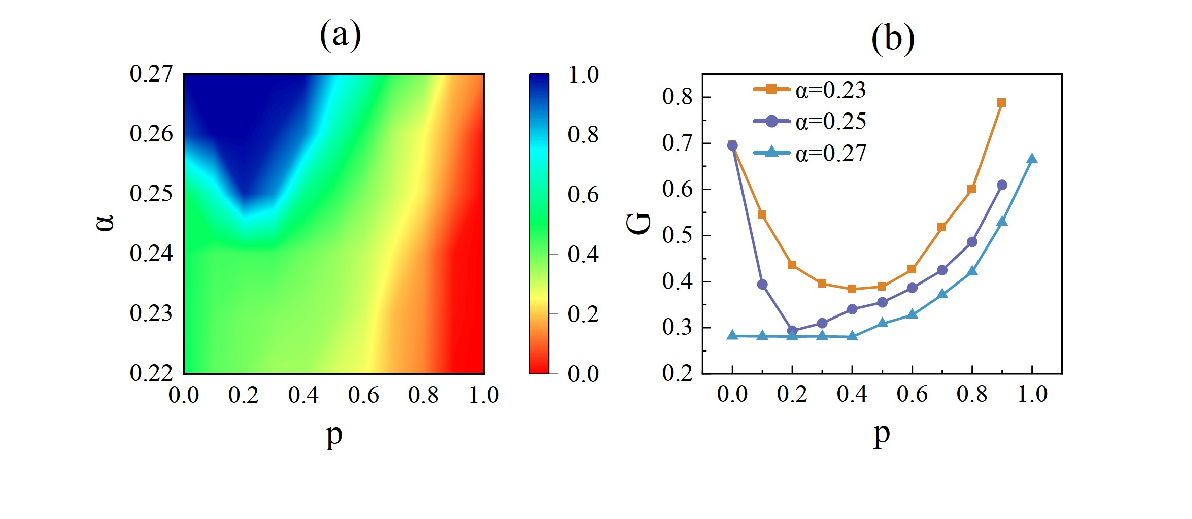}
	\caption{ The diagram of the evolution of cooperation rate and Gini coefficient under the combined effect of external coupling and incentives. (a) $\alpha$-$p$ color-coded plot of cooperation level. (b) Gini coefficient as a function of $p$ for different $\alpha$.\ $r=2.5$ for all panels.
		\label{Fig:fig2} }
\end{figure*}

Our investigation began with an analysis of how the incentive mechanism impacts overall cooperation within the two-layer network. As depicted in Fig.~\ref{Fig:fig2}(a), where cooperation levels are color-coded on the $\alpha$-$p$ parameter plane, we examined the influence of external coupling and incentive strength on cooperative behaviors. Initially, for a constant $p$, incentives effectively stimulated cooperative behaviors across all individuals. However, our analysis revealed three key behaviors of $\alpha$ across different $p$ intervals. At lower $\alpha$ values, no contributions were observed in the two-layer network under external coupling. As $\alpha$ increases within certain intervals, cooperation rate curves exhibit a non-monotonic trend with increasing $p$. For instance, at $\alpha=0.25$, the cooperation rates peaked at $p=0.2$, leading to a substantial enhancement in cooperation within this specific range. Beyond $p=0.5$, however, further increasing of $p$ inhibited mutual assistance, inducing the dominance of defectors and accelerating their infiltration of the two-layer network. This highlights the crucial need to maintain an optimal level of external coupling strength to foster cooperation. Excessive coupling between layers can paradoxically impede cooperative efforts, turning external coupling into a double-edged sword. In addition, although higher value of $\alpha$ expanded the effective scope of cooperation, implementing these collaborations in an unregulated environment poses significant challenges. Therefore, effectively promoting cooperation in two-layer networks depends on achieving a subtle balance between external coupling strength and redistribution synergy effects. We also conducted simulations using different values of $b$ and $r$ and obtained similar results, emphasizing the importance of these factors in shaping collaborative dynamics.

We further analyzed wealth disparities using the Gini coefficient ($G$), a robust measure of wealth concentration. Fig.~\ref{Fig:fig2}(b) presents the Gini coefficient as a function of $p$ for each of the three $\alpha$ values chosen to correspond to the three principal characteristics in panel~(a). It is important to note that the study of inequality in this paper excludes the case of $f_C=0$, because all agents were defectors who contributed nothing and did not receive payoffs, rendering the Gini coefficient meaningless. It was observed that for all $p$ values, an increase in incentive strength also generally reduced the Gini coefficient, i.e., higher incentive strength more severely penalized defector free-riding behavior, which increased the benefits to low-income cooperators and thus reduced overall wealth disparities. Moreover, as $p$ increased, $\alpha$ delineated two distinct behaviors. When $\alpha=0.23$ and $\alpha=0.25$, the Gini coefficient initially decreased and then increased with $p$ across all $\alpha$ parameters, suggesting that external coupling contributed to diminishing wealth disparities in the two-layer network. Conversely, when $\alpha=0.27$, the Gini coefficient remained low before decreasing as $p$ increased. In conclusion, the incentive mechanism served as a significant disincentive to curb inequality between the rich and the poor. Comparing panels~(a) and (b), it was evident that even without an increase in the cooperation rate, higher $p$ values contributed to reducing the Gini coefficient. However, the reduction in wealth inequality was most pronounced within the $\alpha$ range corresponding to the nonlinear increase in cooperation rates, highlighting the critical role of incentives for cooperative behavior in mitigating wealth disparities in two-layer networks.

\begin{figure*}
	\centering
    \includegraphics[width=.8\textwidth]{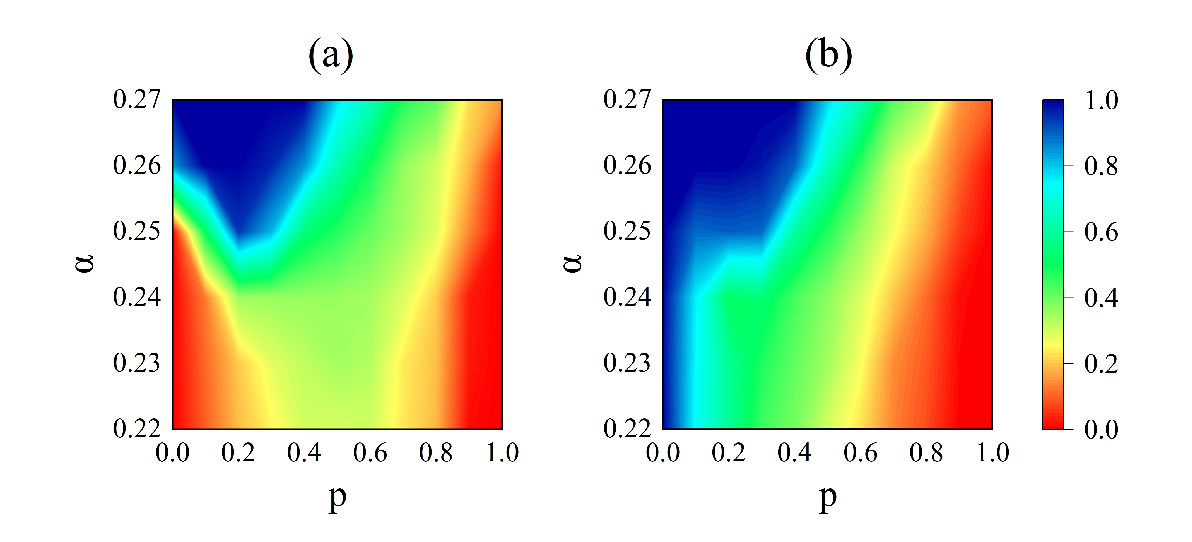}
	\caption{Cooperation level on $\alpha-p$ parameter plane for different layers of the network. Panel~(a) depicts the portion of cooperators on the regular network (a) while panel~(b) shows the same value on the scale-free graph. $r=2.5$ for both panels.
		\label{Fig:fig3} }
\end{figure*}

In addition to strong incentive measures to promote the increase of cooperation rate, moderate coupling strength also played a crucial role, as the cooperative mutual assistance behavior of the two networks reached its strongest under moderate coupling. Therefore, the diversification of structure and strategic interactions in dual-layer networks was crucial for the efficacy of mechanisms. To further dissect the evolutionary impacts of these two parameters on cooperation within heterogeneous two-layer networks, we proceeded to study each network separately. Fig.~\ref{Fig:fig3} shows the $\alpha$-$p$ color-coded plot of cooperation level on the regular and scale-free networks. In panel (a), as $p$ increased, an optimal value emerged that maximized the cooperation rate before it began to decline. Up to $\alpha=0.25$, there was a significant promotion of peak cooperation, even at low $p$ values, emphasizing the regular network as the primary driver of the non-monotonic changes in overall cooperation rates. On the other hand, in panel (b), the scale-free network initially exhibited a complete dominance of cooperators ($f_{C}=1$) when $p=0$, but as $p$ increased, the cooperation rate decreased linearly. Whereas before the cooperation rate in the regular network grew to its peak, the rate of decline in cooperation slowed in the scale-free network. Beyond this point, strategy convergence accelerated, leading to a breakdown in cooperation. External coupling played a crucial positive role in enhancing mutual assistance within the regular network, but it did not facilitate cooperation in the scale-free network.

\begin{figure*}
	\centering
    \includegraphics[width=.8\textwidth]{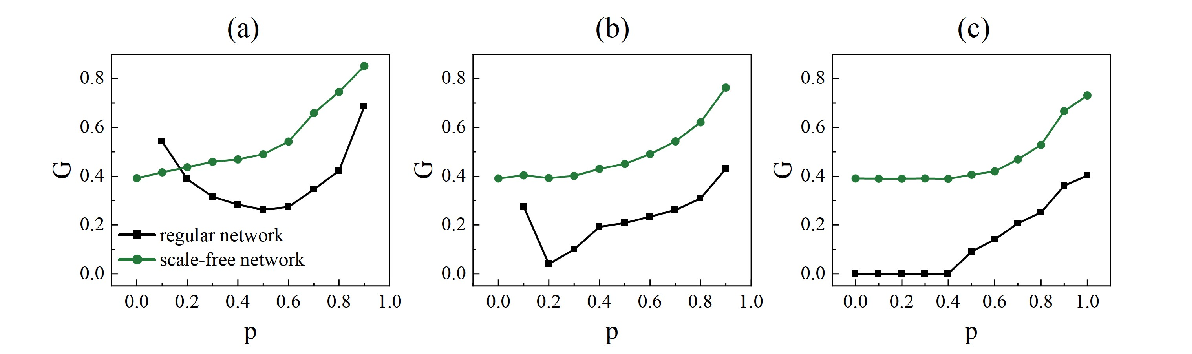}
	\caption{Gini coefficient on the dependence of external coupling for each layer of the network. Panels show $G$ value for $\alpha=0.23$~(a), $\alpha=0.25$~(b), and $\alpha=0.27$~(c). $r=2.5$ for all panels.
		\label{Fig:fig4} }
\end{figure*}

Fig.~\ref{Fig:fig4} illustrates the $p$-dependence of the Gini coefficient on the regular and scale-free networks for three values of $\alpha$ also used in Fig.~\ref{Fig:fig2}(b). Remarkably, the Gini coefficient changed the same dominant theme on the regular network as it did on the two-layer network. Specifically, as shown in panels~(a) and (b), the economic disparity below an $\alpha$-threshold continued to narrow and then rise as $p$ increased, the trend being more pronounced close to the threshold. In comparison, the Gini coefficient in the scale-free network changed less dramatically and fluctuated significantly only when the Gini coefficient in the regular network began to decline, thus exacerbating the negative impact of wealth disparities. In addition, the global Gini coefficient of scale-free networks was generally higher than those in regular networks. This was owing to the fact that unlike regular networks, where each agent had a constant number of neighbors ($k=4$), scale-free networks were characterized by a small number of agents having a large number of neighbors, while a large group of agents had only a few neighbors. While the strength of coupling in scale-free networks failed to bring a significant positive effect, the contribution of the regular networks in a two-layer network resulted in a generally beneficial effect. One conclusion drawn was that the heterogeneity of network information was also a key factor, showing that a certain degree of external coupling could further facilitate the narrowing of the gap between rich and poor in the overall two-layer network system.

To further verify the correlation between cooperation and wealth disparity, we analyzed and summarized the confluence of Fig.~\ref{Fig:fig3} and Fig.~\ref{Fig:fig4}. In both regular and scale-free networks, the Gini coefficient decreased as the cooperation rate rose. There seemed to be a negative correlation between $f_{C}$ and $G$ in single-layer networks. Furthermore, the wealth disparity in the regular network began to decrease significantly as the cooperation rate between the two networks approached. These results further validated the findings in Fig.~\ref{Fig:fig2}(b), suggesting that a moderate strength of external coupling could reduce the disparity by boosting the cooperation rate. Considering the complex and diverse nature of networks that reflected the real world, enhancing partnerships could indeed foster shared prosperity.

As discussed, it was clear that external couplings in a two-layer network augment the channels for obtaining information about invisible neighbors, which exerted both positive and negative impacts on the propagation of cooperative behaviors within the group, especially valid for the regular network. Moreover, the incentive strength catalyzed collective action in the environment brought from a two-layer network with external couplings when it grew to a certain value. In the following, a fixed $\alpha=0.25$ value was chosen to analyze the distribution of agents' strategies in separate networks. A microscopic standpoint offered valuable insights into how the external couplings influenced the networks separately at different strengths. Since the main reason for promoting cooperation was due to external couplings dominated by regular networks, it was useful to examine its structure depicted in Fig.~\ref{Fig:fig5}, focusing on the evolution of the four attribute agents in spatially distributed snapshots under different $p$ values.

\begin{figure*}
	\centering
    \includegraphics[width=.9\textwidth]{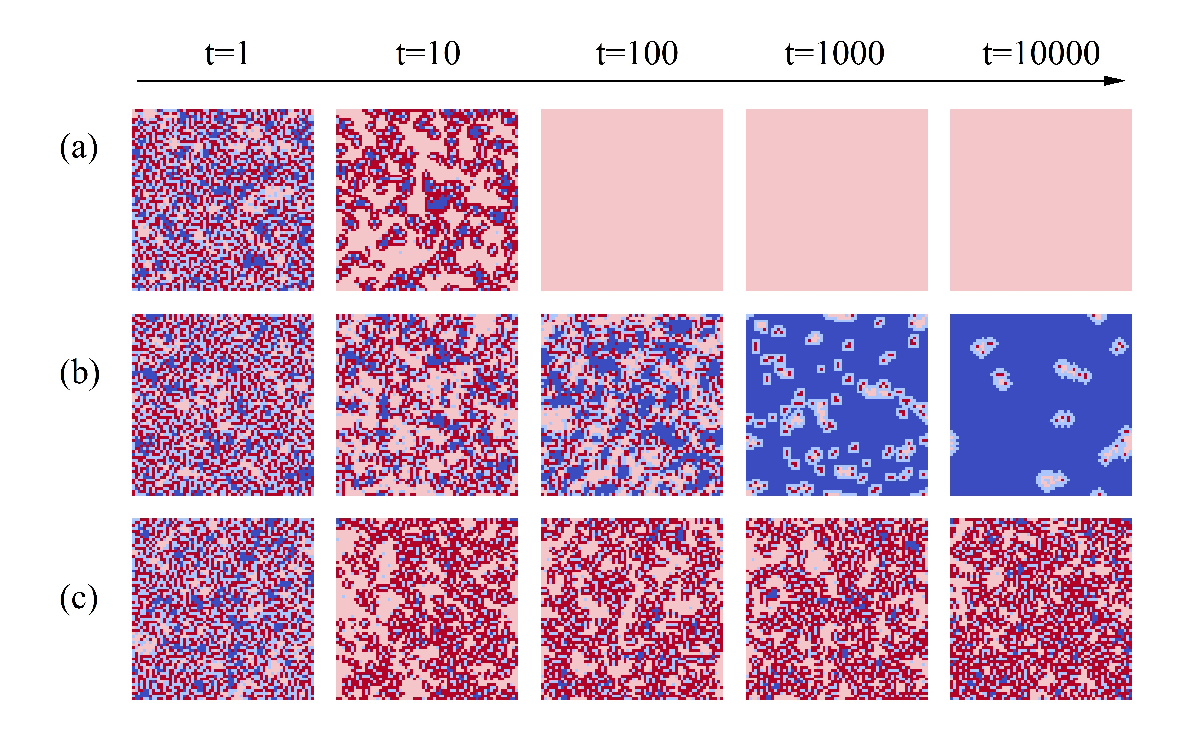}
	\caption{ A micro perspective on the cooperation of agents on the regular network layer. Snapshots of the spatial distribution evolving over time. Rows from top to bottom: $p=0$, $p=0.2$, $p=0.8$, and columns from left to right: $t=1$, $t=10$, $t=100$, $t=1000$, $t=10000$. Initially, each agent was randomly distributed in the network. When the first step was over, the agents formed four different attributes depending on their strategies and benefits: low-income cooperators (light blue), low-income defectors (light red), high-income cooperators (dark blue), and high-income defectors (dark red). The common parameters used in these experiments are $r=2.5$ and $\alpha=0.25$.
		\label{Fig:fig5} }
\end{figure*}

Upon vertical observation of Fig.~\ref{Fig:fig5}, the snapshots at each moment $t$ aligned with the results in Fig.~\ref{Fig:fig3}, as the percentage of cooperators grew and then dropped as $p$ increased as evolution proceeded. Before the start of evolution, cooperators and defectors were randomly located in the regular network. As the game unfolded, small clusters of high-income cooperators and low-income defectors began to emerge at each of the three $p$ values. Evolving to $t=10$ at the early stage of the game, akin to the traditional simulation values, defectors were the initial invaders of the network. Interestingly, the mainstay of the defectors was mostly the low-income people. And cooperators formed denser clusters to defend against the invasion of defectors. At the peripheries of these clusters, cooperators fell victim to exploitation by defectors, leading them to transition into high-income defectors versus low-income cooperators and gradually revealing cross-like structures ~\cite{HauertN2004}. It was worth mentioning that since the state of exploitation and degradation experienced by high-income defectors and low-income cooperators, cross-like structures were unable to extend to tight clusters, but only occurred at the peripheries of different strategy clusters.

As the evolutionary process unfolded, the dynamics in the regular networks at different values of $p$ became increasingly distinct. At $p=0$ and $p=0.8$, the cooperators could not defend against the infiltration of defectors. This was due to the high proportion of high-income defectors under the cross-like structures, who, even with the implementation of incentive mechanisms, were exploited for a much higher amount than the low-income cooperators after rewards. Thankfully, at $p=0.8$, agents were continuously influenced by additional sources of information. While most agents faced increased susceptibility to opportunistic temptations, a small subset successfully maintained cooperation, particularly evident in instances where $p=0.2$, resulting in flourishing cooperative behavior. Moderate coupling strength mitigated wealth disparities in the system, controlling for the number of early high-income defectors. The lower Gini coefficient put the incentive mechanism into play, where the benefits of the rewarded low-income cooperators rivaled that of the defectors under penalization, making them less likely to learn strategies from defectors within the network. In addition, high-income cooperators who formed small clusters in the early period had a stable space in which to release kindness to their neighbors, allowing agents engaged in cooperative behavior to gradually spread to occupy almost the entire network. A combination of coupling strength and incentive mechanism was therefore necessary to maintain fairness in the regular network and to promote the diffusion of cooperative behavior.

\begin{figure*}
	\centering
    \includegraphics[width=.7\textwidth]{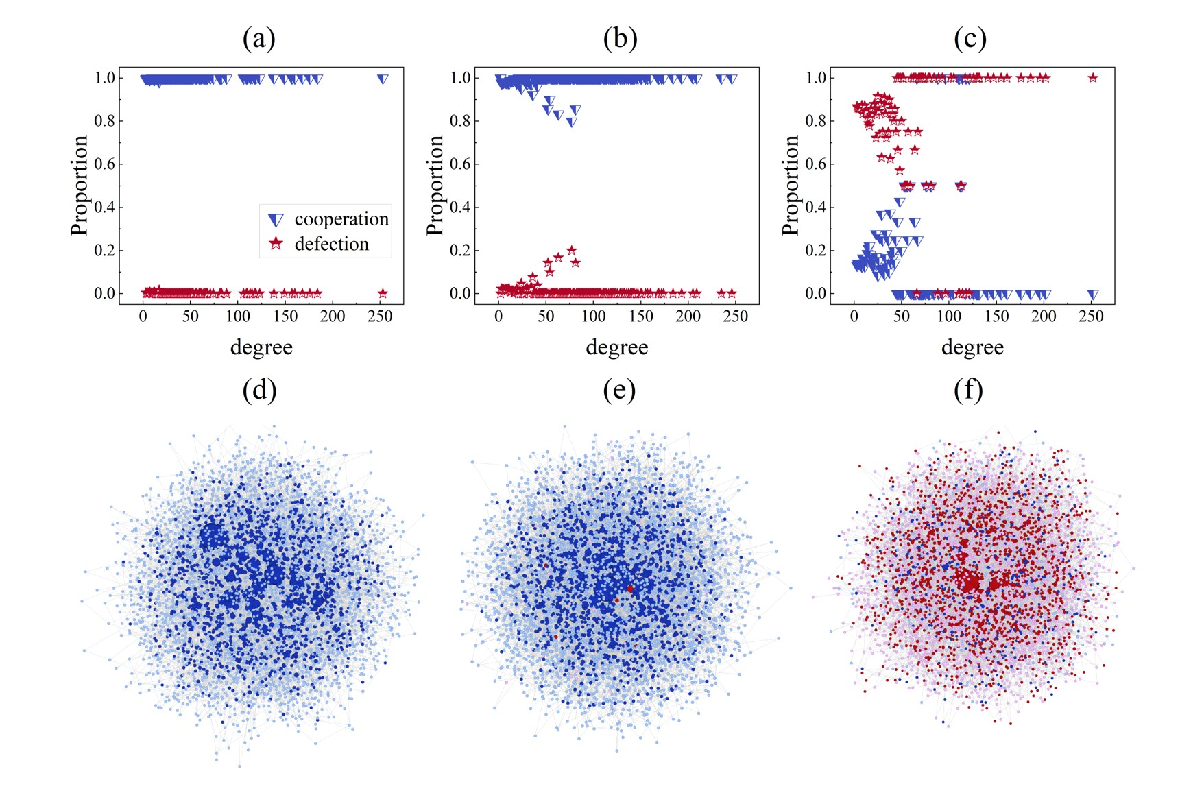}
	\caption{ A micro perspective of cooperation on the scale-free network. The portion of cooperator and defector players in dependence of degree for three representative $p$ values. Panels~(a)-(c) depict the distributions at $p=0$, $p=0.2$, and $p=0.8$, respectively. Panels~(d)-(f) show the spatial distribution of four classes of agents. In agreement with Fig.~\label{Fig:fig5} low-income cooperators are marked by (light blue), low-income defectors by light red, high-income cooperators by dark blue, and high-income defectors by dark red. Furthermore, the size of a node characterizes its degree. The parameters used in the experiments are $r=2.5$ and $\alpha=0.25$.
		\label{Fig:fig6} }
\end{figure*}

\begin{figure*}
	\centering
		 \includegraphics[width=.6\textwidth]{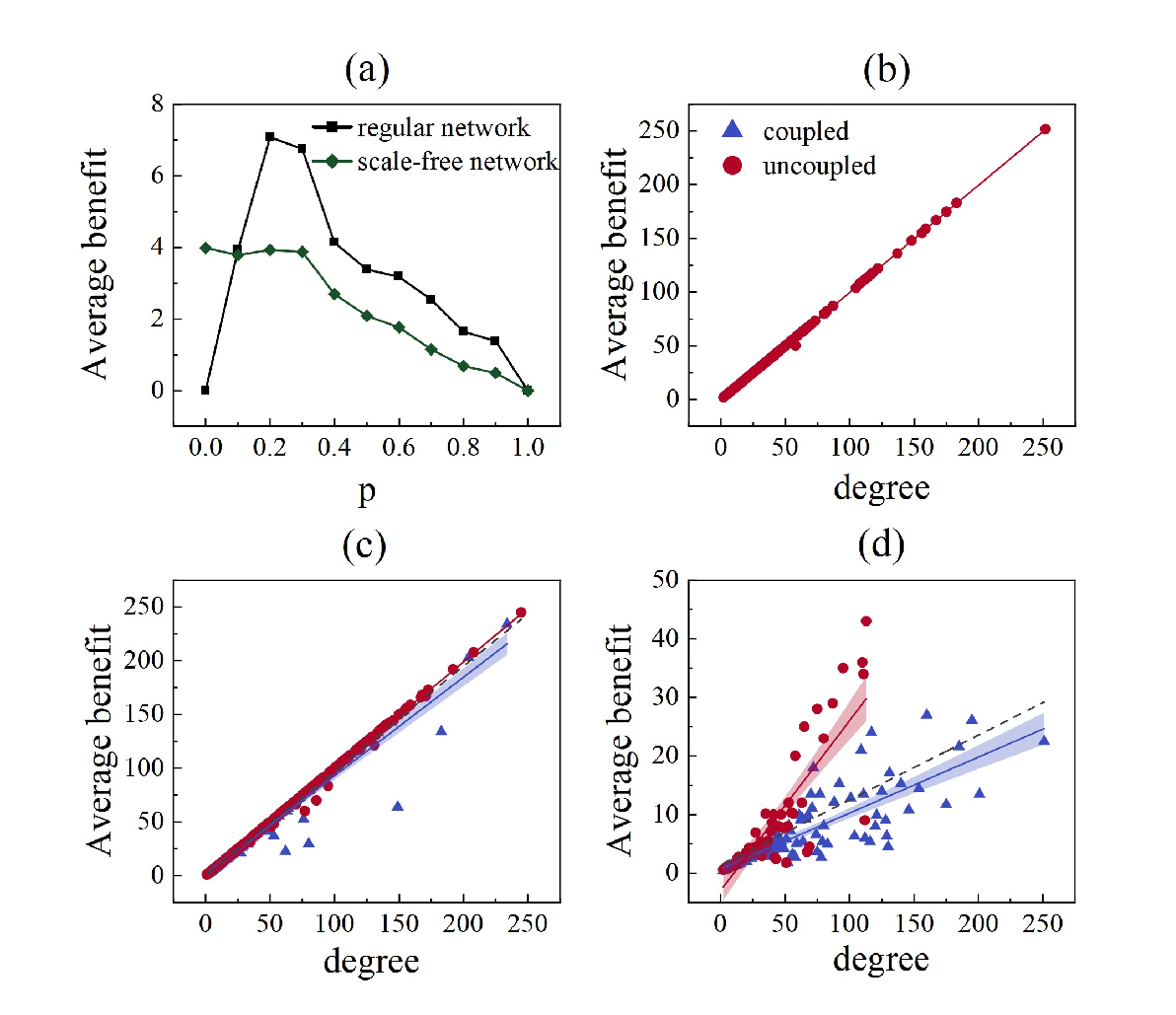}
	\caption{ A micro perspective on average benefits of agents whether or not having external coupling. (a) Average benefit of two networks with different coupling strengths. (b-d) The average wealth distribution in scale-free networks depends on degrees under $p=0$, $p=0.2$, $p=0.8$, respectively. The blue and red nodes represent the average benefit in different degrees of agents who have external coupling or not. Solid blue and red lines represent linear fitting, while blue shading and red shading represent confidence intervals. The dark gray dashed line is fit to the average wealth of all participants in the network at mixed degrees. The parameters used in the experimental results are $r=2.5$ and $\alpha=0.25$.
		\label{Fig:fig7} }
\end{figure*}

The next question was why coupling poses challenges in fostering positive effects in scale-free networks. To understand it more deeply in Fig.~\ref{Fig:fig6}, we presented the distribution of strategies on nodes with different degrees. Panels~(a)-(c) illustrated that as the parameter $p$ increased, the overall proportion of cooperative strategies decreased across different node degrees, while the polarization between the two strategies gradually diminished. Initially, the changes in strategy proportion primarily occurred among nodes with lower degrees (25 to 100). When $p$ reached $0.8$, the increased information sharing intensified the allure of defection, prompting more agents to defect and exploit benefits. Consequently, cooperators were predominantly found among nodes with low to moderate degrees, especially in the latter class. Agents corresponding to this mixed distribution of cooperation and defection across degrees exhibited more fluctuation in strategy learning. Utilizing the same categorization technique, we classified agents into four distinct attributes based on their strategy and payoff. This classification allowed us to observe, in detail, how the value of $p$ influenced different groups and, consequently, affected cooperation rates as shown in the distributions at three representative coupling levels in panels~(a)-(c). Lower coupling strength allows strategy transition to defection among some low-degree nodes only. Given agents' inclination to mimic strategies of larger nodes, the network retains a high level of cooperation. Conversely, higher coupling strength severely impacts high-degree nodes, converting them entirely to defectors, while partial cooperation still exists among nodes with low to moderate degrees. We suspect that the phenomena observed in Fig.~\ref{Fig:fig6} stem from insufficient penalties against affluent defectors. The intricate interplay among nodes in scale-free networks complicates assessing wealth impact through spatial distribution maps. Hence, in Fig.~\ref{Fig:fig7}, we statistically analyze the average wealth of agents subjected to coupling across different degrees to delve deeper into this issue.

To examine the potential impacts of coupling strength in scale-free networks, Fig.~\ref{Fig:fig7} depicts the average benefit of agents with and without external coupling. Panel~(a) extended from average returns to explore their interaction with cooperation. Without and at low coupling strengths, the average benefits in scale-free networks significantly surpassed those in regular networks. As $p$ increased, however, coupled agents adopted strategies of the one with the maximum benefit in adjacent networks, thereby promoting cooperation in regular networks. Conversely, in scale-free networks, coupled agents typically resulted in suboptimal outcomes when acquiring welfare information, gradually falling behind in the cooperation rate. In this scenario, the multigames mechanism caused average returns in regular networks with converging cooperation rates to eventually exceed those in scale-free networks. Beyond a certain threshold of coupling strength, overly coupled agents learned defection strategies from benefit datasets, undermining cooperation in scale-free networks. For instance, high-degree nodes in scale-free networks, once adopting defection strategies, tempted their uncoupled neighbors into similar betrayals. Similarly, adverse effects were observed in regular networks when $p>0.3$, where cooperation rates and average benefits rapidly declined in both networks.

We closely observed the average benefit of agents with mixed degrees in scale-free networks, analyzing the behaviors that resulted in suboptimal outcomes. Panel~(b) of Fig.~\ref{Fig:fig7} showed that in the single-layer scale-free network, the average benefit nearly followed a linear fit. As the coupling $p$ increased, the benefit of some agents significantly declined. This was because these players initially achieved high payoffs through defection in early games, but as evolution progressed, their neighbors converged in strategy due to the allure of their high payoffs, turning the advantage of coupled agents into a disadvantage within the defecting group, making it difficult for them to make substantial profits. As shown in Fig.~\ref{Fig:fig6}(b), some nodes with lower degrees were the first to engage in defection. The decrease in the average benefits of certain agents also indicates a misalignment between cooperative objectives and the maximization of social welfare~\cite{T. A. Han2024}. The phenomenon of lower benefits for coupled agents compared to uncoupled ones due to defection was particularly evident at $p=0.8$. In conjunction with Fig.~\ref{Fig:fig6}(c), high-degree coupled agents were predominantly defectors, deeply entrenched in their benefits dataset, resistant to mindset changes that would prompt strategy adjustments even though they faced penalties. In contrast, the benefits of nodes with medium to low to moderate degrees fluctuated around the fitted line, indicating they employed diversified strategies. Additionally, panels~(b)-(d) of Fig.~\ref{Fig:fig6} emphasized that the increase in the Gini coefficient in scale-free networks primarily stemmed from the widening average benefits gap between coupled and uncoupled agents, and the mixed strategies adopted by coupled agents with low to moderate degrees. Therefore, the incentive mechanism in scale-free networks primarily generated positive effects internally, while increased coupling strength resulted in suboptimal outcomes, thus inhibiting these incentive effects.

\section{Conclusion}
Cooperation is a fundamental element of human society, and the diversity of individual strategies leads to wealth inequality and reduces prosocial behavior among vulnerable groups. 
To avoid this undesired outcome we here proposed a wealth redistribution incentive mechanism in combination with a coupling that connects agents across different networks as acquaintances, thereby facilitating access to shared datasets.
Through extensive simulation experiments, we identify an incentive threshold where external coupling above this threshold effectively enhances cooperation rates, with regular networks playing a pivotal enabler. Our focus lies in exploring the non-monotonic growth range of cooperation, revealing its intricate relationship with wealth inequality. In single-layer networks, we observe a negative correlation between cooperation rates and the Gini coefficient, providing novel insights. From a micro perspective, moderate external coupling strength within regular networks facilitates mitigating wealth disparities, and lower Gini coefficients amplify the effectiveness of incentives, thereby maintaining the wealth status of cooperators and reinforcing their strategic choices. In contrast, in scale-free networks, incentive primarily favors within-network effects, whereas increased external coupling strength may lead to suboptimal outcomes, suppressing these incentive effects. Hence, precise control of incentive and external coupling strength through optimized combinations effectively upholds fairness within regular networks and promotes the diffusion of cooperative behaviors.

The combined effect of an incentive mechanism and external coupling has established a thriving multi-layer network society where a harmonious community can sustain high levels of cooperation while keeping Gini coefficients low. The comprehensive consideration of cooperation and wealth inequality reveals their inseparable connection, providing insights into the proportional collection of personal income tax and improving overall population satisfaction by taxing non-collaborators. In addition, an appropriate proportion of long-term connections between different individuals, institutions, or countries can promote universal cooperation and prosperity, especially in supporting vulnerable groups. However, if one party is at a disadvantage, appropriate external coupling strength can effectively reduce losses. It is worth noting that deferred incentives were not considered in this study, as in practical situations, the collection of interest information within organizations is usually not completed from the beginning, and the reward and penalty mechanisms are usually carried out after performance evaluation.

\section*{Acknowledgements}
The research for this work was supported by Major Humanities and Social Sciences Research Projects in Zhejiang higher education institutions (Grant No.~2023QN109) and by the National Research, Development and Innovation Office under Grant No.~K142948.

\section*{References}

\providecommand{\newblock}{}

\end{document}